\documentclass[10pt]{article}
\usepackage{subfigure}

\newcommand\arcmin{\mbox{$^\prime$}}%
\newcommand\phn{\phantom{0}}%
\usepackage{lineno}
\usepackage{graphics}
\usepackage{graphicx}

\usepackage{scicite}

\usepackage{times}

\newenvironment{sciabstract}{%
\begin{quote} \bf}
{\end{quote}}

\newcounter{lastnote}

\title{Detection of 16 Gamma-Ray Pulsars Through Blind Frequency Searches Using the Fermi LAT}

\author
{A.~A.~Abdo$^{1,2}$ M.~Ackermann$^{3}$ M.~Ajello,$^{3}$ B.~Anderson,$^{4}$ W.~B.~Atwood,$^{4}$ M.~Axelsson,$^{5,6}$ \and 
L.~Baldini,$^{7}$ J.~Ballet,$^{8}$ G.~Barbiellini,$^{9,10}$ M.~G.~Baring,$^{11}$
% ...
\\
\\
\\
\normalsize{$^{\ast}$To whom correspondence should be addressed:}\\
\normalsize{dormody@scipp.ucsc.edu, paul.ray@nrl.navy.mil, pablo@scipp.ucsc.edu, ziegler@scipp.ucsc.edu}\\
}

\date{}

\begin{document}




\maketitle 

\noindent
$^{1}$National Research Council Research Associate\\
$^{2}$Space Science Division, Naval Research Laboratory, Washington, DC 20375\\
$^{3}$W. W. Hansen Experimental Physics Laboratory, Kavli Institute for Particle Astrophysics and Cosmology, Department of Physics and SLAC National Accelerator Laboratory, Stanford University, Stanford, CA 94305\\
$^{4}$Santa Cruz Institute for Particle Physics, Department of Physics and Department of Astronomy and Astrophysics, University of California at Santa Cruz, Santa Cruz, CA 95064\\
$^{5}$The Oskar Klein Centre for Cosmo Particle Physics, AlbaNova, SE-106 91 Stockholm, Sweden\\
$^{6}$Department of Astronomy, Stockholm University, SE-106 91 Stockholm, Sweden\\
$^{7}$Istituto Nazionale di Fisica Nucleare, Sezione di Pisa, I-56127 Pisa, Italy\\
$^{8}$Laboratoire AIM, CEA-IRFU/CNRS/Universit\'e Paris Diderot, Service d'Astrophysique, CEA Saclay, 91191 Gif sur Yvette, France\\
$^{9}$Istituto Nazionale di Fisica Nucleare, Sezione di Trieste, I-34127 Trieste, Italy\\
$^{10}$Dipartimento di Fisica, Universit\`a di Trieste, I-34127 Trieste, Italy\\
$^{11}$Rice University, Department of Physics and Astronomy, MS-108, P. O. Box 1892, Houston, TX 77251, USA\\
$^{12}$Istituto Nazionale di Fisica Nucleare, Sezione di Padova, I-35131 Padova, Italy\\
$^{13}$Dipartimento di Fisica ``G. Galilei", Universit\`a di Padova, I-35131 Padova, Italy\\
$^{14}$Department of Physics, Center for Cosmology and Astro-Particle Physics, The Ohio State University, Columbus, OH 43210\\
$^{15}$Istituto Universitario di Studi Superiori (IUSS), I-27100 Pavia, Italy\\
$^{16}$Istituto Nazionale di Fisica Nucleare, Sezione di Perugia, I-06123 Perugia, Italy\\
$^{17}$Dipartimento di Fisica, Universit\`a degli Studi di Perugia, I-06123 Perugia, Italy\\
$^{18}$Dipartimento di Fisica ``M. Merlin" dell'Universit\`a e del Politecnico di Bari, I-70126 Bari, Italy\\
$^{19}$Istituto Nazionale di Fisica Nucleare, Sezione di Bari, 70126 Bari, Italy\\
$^{20}$Laboratoire Leprince-Ringuet, \'Ecole polytechnique, CNRS/IN2P3, Palaiseau, France\\
$^{21}$Department of Physics, University of Washington, Seattle, WA 98195-1560\\
$^{22}$INAF-Istituto di Astrofisica Spaziale e Fisica Cosmica, I-20133 Milano, Italy\\
$^{23}$NASA Goddard Space Flight Center, Greenbelt, MD 20771\\
$^{24}$George Mason University, Fairfax, VA 22030\\
$^{25}$Laboratoire de Physique Th\'eorique et Astroparticules, Universit\'e Montpellier 2, CNRS/IN2P3, Montpellier, France\\
$^{26}$Department of Physics, Royal Institute of Technology (KTH), AlbaNova, SE-106 91 Stockholm, Sweden\\
$^{27}$Department of Physics, Stockholm University, AlbaNova, SE-106 91 Stockholm, Sweden\\
$^{28}$Royal Swedish Academy of Sciences Research Fellow, funded by a grant from the K. A. Wallenberg Foundation\\
$^{29}$Agenzia Spaziale Italiana (ASI) Science Data Center, I-00044 Frascati (Roma), Italy\\
$^{30}$Dipartimento di Fisica, Universit\`a di Udine and Istituto Nazionale di Fisica Nucleare, Sezione di Trieste, Gruppo Collegato di Udine, I-33100 Udine, Italy\\
$^{31}$CNRS/IN2P3, Centre d'\'Etudes Nucl\'eaires Bordeaux Gradignan, UMR 5797, Gradignan, 33175, France\\
$^{32}$Universit\'e de Bordeaux, Centre d'\'Etudes Nucl\'eaires Bordeaux Gradignan, UMR 5797, Gradignan, 33175, France\\
$^{33}$Department of Physical Sciences, Hiroshima University, Higashi-Hiroshima, Hiroshima 739-8526, Japan\\
$^{34}$University of Maryland, College Park, MD 20742\\
$^{35}$University of Alabama in Huntsville, Huntsville, AL 35899\\
$^{36}$Waseda University, 1-104 Totsukamachi, Shinjuku-ku, Tokyo, 169-8050, Japan\\
$^{37}$Cosmic Radiation Laboratory, Institute of Physical and Chemical Research (RIKEN), Wako, Saitama 351-0198, Japan\\
$^{38}$Department of Physics, Tokyo Institute of Technology, Meguro City, Tokyo 152-8551, Japan\\
$^{39}$Centre d'\'Etude Spatiale des Rayonnements, CNRS/UPS, BP 44346, F-30128 Toulouse Cedex 4, France\\
$^{40}$Istituto Nazionale di Fisica Nucleare, Sezione di Roma ``Tor Vergata", I-00133 Roma, Italy\\
$^{41}$Department of Physics and Astronomy, University of Denver, Denver, CO 80208\\
$^{42}$Max-Planck Institut f\"ur extraterrestrische Physik, 85748 Garching, Germany\\
$^{43}$Sterrenkundig Institut ``Anton Pannekoek", 1098 SJ Amsterdam, Netherlands\\
$^{44}$Institut de Ciencies de l'Espai (IEEC-CSIC), Campus UAB, 08193 Barcelona, Spain\\
$^{45}$Space Sciences Division, NASA Ames Research Center, Moffett Field, CA 94035-1000\\
$^{46}$NYCB Real-Time Computing Inc., Lattingtown, NY 11560-1025\\
$^{47}$Department of Chemistry and Physics, Purdue University Calumet, Hammond, IN 46323-2094\\
$^{48}$Institute of Space and Astronautical Science, JAXA, 3-1-1 Yoshinodai, Sagamihara, Kanagawa 229-8510, Japan\\
$^{49}$Max-Planck-Institut f\"ur Kernphysik, D-69029 Heidelberg, Germany\\
$^{50}$Instituci\'o Catalana de Recerca i Estudis Avan\c{c}ats (ICREA), Barcelona, Spain\\
$^{51}$Consorzio Interuniversitario per la Fisica Spaziale (CIFS), I-10133 Torino, Italy\\
$^{52}$Center for Research and Exploration in Space Science and Technology (CRESST), NASA Goddard Space Flight Center, Greenbelt, MD 20771\\
$^{53}$University of Maryland, Baltimore County, Baltimore, MD 21250\\
$^{54}$Dipartimento di Fisica, Universit\`a di Roma ``Tor Vergata", I-00133 Roma, Italy\\
$^{55}$School of Pure and Applied Natural Sciences, University of Kalmar, SE-391 82 Kalmar, Sweden\\


\begin{sciabstract}
Pulsars are rapidly-rotating, highly-magnetized neutron stars emitting radiation across the electromagnetic spectrum. Although there are more than 1800 known radio pulsars, until recently, only seven were observed to pulse in gamma rays and these were all discovered at other wavelengths. The \emph{Fermi} Large Area Telescope makes it possible to pinpoint neutron stars through their gamma-ray pulsations. We report the detection of 16 gamma-ray pulsars in blind frequency searches using the LAT. Most of these pulsars are coincident with previously unidentified gamma-ray sources, and many are associated with supernova remnants. Direct detection of gamma-ray pulsars enables studies of emission mechanisms, population statistics and the energetics of pulsar wind nebulae and supernova remnants.

\end{sciabstract}


A wide variety of astrophysical phenomena, such as black holes, active galactic nuclei, gamma-ray bursts, and pulsars are known to produce photons exceeding many mega electron volts. Detection and accurate localization of sources at these energies is challenging both because of the low fluxes involved and the limitations of the detection techniques.
The sky above 100 MeV was surveyed more than 30 years ago by the COS-B satellite ({\it 1}), and more recently by EGRET ({\it 2}), on 
the \emph{Compton Gamma-Ray Observatory}. One of the main legacies of EGRET was the detection of $\sim$300 gamma-ray sources, many of which have remained unidentified, despite 
searches at a wide variety of wavelengths ({\it 3}). Many EGRET (and COS-B) unidentified sources are thought to be of Galactic origin, because of
their lack of variability and concentration along the Galactic plane. A large fraction of these have been suspected to be pulsars [e.g. ({\it 4--6})] despite deep radio and X-ray searches often failing to uncover pulsed emission, even when the gamma-ray sources were coincident with supernova remnants (SNR) or pulsar wind nebulae (PWN). The lack of radio pulsations has usually been explained as the narrow radio beams missing the line of sight toward the Earth ({\it 7}). We refer to such pulsars as ``radio-quiet"; even though they may emit radio waves, these cannot be detected at Earth. Before the launch of \emph{Fermi}, Geminga ({\it 8}) was the only known radio-quiet gamma-ray pulsar. Current models of pulsar gamma-ray emission predict that gamma-ray beams are much wider than radio beams ({\it 9}), thus suggesting that there may be a large population of radio-quiet gamma-ray pulsars.

Soon after launch on 11 June 2008, the Large Area Telescope (LAT) on \emph{Fermi} began surveying the sky at energies above 20 MeV. 
A companion paper describes the LAT detection of a population of gamma-ray millisecond pulsars (MSPs) ({\it 10}). Here we report the detection of 16 pulsars found in blind 
frequency searches using the LAT. Previously, gamma-ray pulsars had been detected only by using a radio (or in the case of Geminga, an X-ray) ephemeris. 

\paragraph*{Observations and Data Analysis.}
The LAT is a high-energy gamma-ray telescope sensitive to photon energies from 20 MeV to over 300 GeV, featuring a solid-state silicon tracker, 
a cesium-iodide calorimeter and an anti-coincidence detector ({\it 11}). Gamma-ray events recorded in the LAT have time stamps derived from
a GPS clock on the \emph{Fermi} satellite with an accuracy of $<1 \mu$s ({\it 12}). \emph{Fermi} operates in continuous sky survey mode, covering the entire sky 
every 3 hours. Compared to EGRET, \emph{Fermi} has a larger effective area (9,500 cm$^2$ at normal incidence), larger field of view (2.4 sr), more efficient use of time on orbit for photon collection, and a finer point spread function ($5^{\circ}$ at 100 MeV, $0.8^{\circ}$ at 1 GeV). The first three factors result in more rapid photon accumulation and the fourth increases the signal to noise ratio by improving the background rejection. 

Even with the improvements of the LAT, gamma-ray data remain extremely sparse. The brightest steady gamma-ray source in the sky, the Vela pulsar, spins 11 
times per second; however, it provides fewer than one hundred ($>$ 30 MeV) photons during every two orbits of the LAT ({\it 13}). As a result, detection of gamma-ray pulsations from sources with more typical brightness requires weeks or months of data. A standard technique for finding a periodic signal in a data set is a 
fast Fourier transform (FFT). A fully coherent FFT becomes memory intensive because the number of frequency bins in the FFT increases with the length of the observational time, as 
$N_{\rm{bins}} = 2 T f_{\rm{max}}$, where $T$ is the duration of the observation and $f_{\rm{max}}$ is the maximum frequency. Furthermore, pulsars
gradually spin down as they radiate away energy, requiring the computation of many tens of thousands of FFTs to scan a realistic frequency and frequency derivative ($f$ and $\dot{f}$) 
parameter space, making FFT searches computationally intensive, if not prohibitive [e.g. ({\it 14})]. We thus used the time-differencing technique ({\it 15}) 
and a fixed window of $T = 2^{19}$s ($\sim$1 week). This dramatically reduces the number of FFT bins and frequency derivative trials. This method was successfully applied to EGRET 
data ({\it 16}).

We applied the time-difference technique to photons from a small region of interest around selected target positions. We used two sets of target 
positions. The first was a list of $\sim$100 promising locations derived from multiwavelength observations. This list was selected based on location in the Galactic plane, 
spectra, lack of long-term variability, and presence of an SNR, PWN, or a likely neutron star identified from X-ray 
observations. Many of these locations were unidentified gamma-ray sources from EGRET. The second list, compiled after the LAT started collecting data, included locations of 
gamma-ray sources from a preliminary version of the LAT catalog. After eliminating sources with probable active galactic nuclei associations, we ended up with $\sim200$ locations, of which 36 were consistent with previously unidentified EGRET sources. 

We analyzed data collected from sky survey observations beginning on 4 August 2008 and ending on 25 December 2008. We applied a 
barycenter correction using the JPL DE405 Solar System ephemeris ({\it 17}), assuming that the source position was either a likely candidate counterpart, found through 
multiwavelengths studies, or our best estimate based on LAT data. This translation into
the nearly inertial reference frame of the Solar System Barycenter (SSB) corrects for classical and relativistic effects of the motion of the Earth. We accepted all highest quality 
[``Diffuse'' class ({\it 11})] photons falling within a $0.8^{\circ}$ radius of our source location 
with energies greater than 300 MeV. For each target, we searched a set of trial frequency derivatives from zero to the spin-down of the young Crab pulsar 
($\dot{f} = -3.7 \times 10^{-10}$ Hz s$^{-1}$). For each trial, we corrected the photon arrival times using the assumed frequency derivative, calculated the time 
differences and computed the FFT sampled with a Nyquist frequency of 64 Hz, and searched the spectrum for significant peaks.

We refined candidate signals by performing an epoch-folding search over a narrow region of frequency and frequency-derivative space using PRESTO ({\it 18}). 
Using the parameters from the time differencing search as an initial timing model, we split the data into intervals 
($\sim 2-4$ weeks, depending on signal to noise), and measured pulse times of arrival (TOAs) in each segment. These TOAs were fit to a timing model using the 
Tempo2 pulsar timing package ({\it 19}). We selected the reference epoch for all timing models to be MJD 54754.0, roughly in the middle of the
observation interval. This choice minimized the covariances between the timing model parameters. We identified 16 pulsars within the first 4 months of data, and 
confirmed these using an independent data set of at least 1 additional month (Figs. 1 and 2; Tables 1 and 2).

Six of the 16 pulsars were found using the positions of well-localized counterparts at other wavelengths, whereas 10 were found using LAT source positions. The LAT positions 
were uncertain by several arcminutes, and in many cases the initial timing models showed very large non-white residuals indicative of a significant distance between the true 
position and the position being used to barycenter the data. We refined the positions in several ways (see below). The number of digits used in the declination of the names assigned to 
the pulsars reflects our confidence in their localization. The discovery frequency may not be the true pulsar rotation frequency, but rather a harmonic or a subharmonic, especially for 
gamma-ray pulse profiles with two peaks separated by $\sim$180 degrees in phase. We tested for this effect, but for certain cases (e.g. J0357+32) the results remain inconclusive, because
of the low photon counts.


\paragraph*{The gamma-ray pulsars.} The 16 pulsars are representative of the highest spin-down luminosity portion of the general population, which includes 1800 pulsars in 
the ATNF database ({\it 20}), including the 6 gamma-ray pulsars detected with EGRET, as well as 8 gamma-ray MSPs detected by $Fermi$-LAT ({\it 10}). 


Thirteen LAT pulsars are associated with unidentified EGRET sources. Indeed, 15 out of 36 EGRET unidentified sources searched showed pulsations, but two of them 
(PSR J1028-5819 and PSR J2021+3651) are known radio pulsars ({\it 21, 22}). Three pulsars (J0357+32, J1459$-$60, and J2238+59) correspond to newly seen LAT sources. This result suggests that the blind searches are flux limited; although many LAT unidentified sources also could be pulsars, most have too low a flux for a pulsation to have been detected in our available data. Five pulsars are likely associated with PWN and/or SNRs and an additional one, J1836+5925, is associated with a known isolated neutron star. J0007+7303 ({\it 23}), which was long suspected of being a pulsar because of its clear association with SNR CTA 1 containing a PWN. J1418-6058 is in the complex Kookaburra region of the Galactic plane and is likely associated with PWN G313.3+0.1, the ``Rabbit'' ({\it 24}). J1809-2332, likely powering the ``Taz'' PWN, is also possibly associated with a recently-discovered mixed-morphology type SNR, G7.5--1.7({\it 25}) . J1826-1256 is probably powering the ``Eel" PWN ({\it 26}), whereas J2021+4026 reveals that a pulsar is present, as long suspected, inside the gamma Cygni SNR ({\it 27}). Two more pulsars present plausible associations with SNR: J0633+0632, in the fairly old ($\sim30$ kyr) complex Monoceros Loop SNR (G205.5+0.5) at a distance of $\sim1.5$ kpc ({\it 28}). J1907+0601 is $\sim0.3^{\circ}$ from SNR G40.5--0.5, but additional evidence is needed before an association can be demonstrated. J1836+5925 is the long sought pulsar powering 3EG 1835+5918, the brightest and most accurately positioned of the unidentified EGRET sources ({\it 29}). Our pulsar coincides with RX J1836+5925, an isolated neutron star for which X-ray pulsations were extensively searched, but not found ({\it 30}). Deep radio pulsation searches with the NRAO Green Bank Telescope (GBT) placed an upper limit at 1.4 GHz on flux density of 7 $\mu$Jy for P$\geq$10 ms ({\it 30}). The X-ray counterpart of J1835+5918 was already suspected to be similar to Geminga, just a bit older and slightly farther away  (within $\sim$800 pc ({\it 31})). Using $\sim130$ ks of publicly available \emph{Chandra} HRC X-ray observations of RX J1836.2+5925 (the same data searched by ({\it 30})), we searched for X-ray pulsations using our gamma-ray ephemeris, but found none. In addition to J1836+5925, three other pulsars are located off the plane, at $|b| > 3^{\circ}$ (J0007+7303, J0357+32, J1741-2054), whereas the remaining 12 are located in the Galactic plane. 

We imaged the error boxes of nine pulsars with specifically targeted short ($\sim$5 ks) Swift observations, and in four cases identified X-ray point sources within the LAT error circle that may be the X-ray counterparts (J0633+0632, J1741-2054, J1813-1246, and J1958+2846). In one other case, J2032+4127, \emph{Chandra} observations ({\it 32}) have revealed a number of X-ray point sources in the region. The best timing solution was found assuming the position of MT91 221, but this cannot be the true X-ray counterpart as it is known to be a B star. The region around J1418-6058 was mapped in detail by \emph{Chandra} ({\it 24}), and a promising point source, denoted R1, is consistent with the LAT source position and likely the X-ray counterpart to the pulsar. \emph{Chandra} observations of 3EG 2020+4017 show several possible counterparts to J2021+4026 of which S21 ({\it 33}) has the best timing solution and is also most consistent with the LAT source position. Other pulsars were better localized using grid searches (J1459-60 and J1732-31) or on-pulse photon analyses (J0357+32, J1907+0601, and J2238+59). In our grid searches, we scanned 25 equally-spaced locations overlapping the 95\% LAT error circle, and looked for the position with the best timing solution. In on-pulse photon analyses, we used only the on-pulse photons to improve the source position.

Five pulsars suffer from large ($\sim0.1^\circ$) positional uncertainties, which degrade the frequency derivative in the timing solution. The problem is particularly acute in 
J0357+32 (Table 2), and the $\sim5\arcmin$ uncertainty in its position results in an uncertainty in the true value of the frequency derivative of $\sim6\times10^{-14}$. Any parameters 
derived from its timing solution are not reliable.

\paragraph*{Implications.} Data from EGRET, COS-B, and SAS-2 demonstrated that rotation-powered 
pulsars are persistent sources of pulsed gamma rays. EGRET results also showed that the luminosity of detected pulsars over all wavelengths was dominated by the gamma-ray part of the spectrum. 
EGRET further established a list of persistent sources at low and intermediate Galactic latitudes that could not be demonstrated to show pulsations. The $Fermi$-LAT data confirms most of 
these sources, and shows that many of these persistent Galactic gamma-ray sources are pulsars. Of 36 LAT sources with EGRET counterparts that were searched, 15 show pulsations (13 from previously 
unknown pulsars).

Until the launch of \emph{Fermi}, the only known radio-quiet pulsar was Geminga, at a distance of $\sim$250 pc ({\it 34}), and with a spin-down luminosity of 
3.2$\times10^{34}$ erg s$^{-1}$ and a gamma-ray efficiency of a few percent (for a 1 sr beam) or 50\% (for isotropic emission) ({\it 8}). It was already apparent 
$\sim30$ years ago, based on COS-B observations ({\it 1}) that the average unidentified gamma-ray source should have a luminosity greater 
than $10^{35}$ erg s$^{-1}$ to be consistent with the very narrow distribution of low latitude gamma-ray sources ({\it 4, 35}). Later studies with EGRET confirmed this ({\it 5}), showing that most unidentified sources were at $\sim$1.2--6 kpc, and had luminosities of $0.7\times 10^{35}$ to $16.7\times 10^{35}$ erg s$^{-1}$. Thus, most unidentified sources could not be like Geminga [e.g.({\it 8})]. The detection, by the LAT, of mostly high-luminosity, low Galactic latitude, and probably relatively distant pulsars confirms the predictions that the unidentified gamma-ray sources could not all be nearby, low-luminosity pulsars such as Geminga.

This new population of gamma-ray selected pulsars helps reveal the geometry of emission from rotation-powered pulsars. A broad gamma-ray beam is required for at least part of our population of pulsars, as we see the high-energy beam but not the relatively narrow radio beam. Thus, this population favors pulsar emission models in which the high-energy radiation occurs in the outer magnetosphere, nearer to the light cylinder. Outer gap models [e.g. ({\it 6})], where particles are accelerated and radiate in vacuum gaps in the outer magnetosphere, and slot gap models [e.g. ({\it 36})], where particles are accelerated and radiate at all altitudes along the open magnetic field boundary, both can reproduce the wide, double-peaked light curves we observed ({\it 9}).

\subsection*{References and Notes}
\begin{itemize}
\item[1.]
B.~N. Swanenburg, {\it et al.}, {\it ApJ} {\bf 243}, L69 (1981).
\item[2.]
R.~C. Hartman, {\it et al.}, {\it ApJS} {\bf 123}, 79 (1999).
\item[3.]
D.~J. Thompson, {\it Reports on Progress in Physics} {\bf 71}, 116901 (2008).
\item[4.]
D.~J. Helfand, {\it MNRAS} {\bf 267}, 490 (1994).
\item[5.]
R. Mukherjee, {\it et al.}, {\it ApJ} {\bf 441}, L61 (1995).
\item[6.]
I.-A. Yadigaroglu, R.~W. Romani, {\it ApJ} {\bf 449}, 211 (1995).
\item[7.]
K.~T.~S. Brazier, S. Johnston, {\it MNRAS} {\bf 305}, 671 (1999).
\item[8.]
G.~F. Bignami, P.~A. Caraveo, {\it ARA\&A} {\bf 34}, 331 (1996).
\item[9.]
K.~P. Watters, R.~W. Romani, P. Weltevrede, S. Johnston, {\it ApJ} {\bf 695}, 1289 (2009).
\item[10.]
A.~A. Abdo, {\it et al.}, ``A Population of Gamma-Ray MSPs seen with the Fermi LAT", {\it Science} (2009)
\item[11.]
W.~B. Atwood, {\it et al.}, {\it ApJ} {\bf 697}, 1071 (2009).
\item[12.]
A.~A. Abdo, {\it et al.}, ``The On-orbit Calibrations for the Fermi Large Area Telescope", {\it Astroparticle Physics} submitted, arXiv {\tt 0904.2226}  (2009). 
\item[13.]
A.~A. Abdo, {\it et al.}, {\it ApJ} {\bf 696}, 1084 (2009).
\item[14.]
A.~M. Chandler, {\it et al.}, {\it ApJ} {\bf 556}, 59 (2001).
\item[15.]
W.~B. Atwood, M. Ziegler, R.~P. Johnson, B.~M. Baughman, {\it ApJ} {\bf 652}, L49 (2006).
\item[16.]
M. Ziegler, B.~M. Baughman, R.~P. Johnson, W.~B. Atwood, {\it ApJ} {\bf 680}, 620 (2008).
\item[17.]
E.~M. Standish, {\it A\&A} {\bf 336}, 381 (1998).
\item[18.]
S.~M. Ransom, {\it Harvard University}, PhD Thesis, (2001).
\item[19.]
G. Hobbs, R. Edwards, R. Manchester, {\it Chinese Journal of Astronomy and Astrophysics Supplement}, {\bf 6}, 020000 (2006)
\item[20.]
R.~N. Manchester, G.~B. Hobbs, A. Teoh, M. Hobbs, {\it AJ} {\bf 129}, 1993 (2005)
\item[21.]
A.~A. Abdo, {\it et al.}, {\it ApJ} {\bf 695}, L72 (2009).
\item[22.]
A.~A. Abdo, {\it et al.}, ``Pulsed Gamma-rays from PSR J2021+3651 with the Fermi Large Area Telescope", {\it ApJ} submitted, arXiv {\tt 0905.4400} (2009).
\item[23.]
A.~A. Abdo, {\it et al.}, {\it Science} {\bf 322}, 1218 (2008).
\item[24.]
C.-Y. Ng, M.~S.~E Roberts, R.~W. Romani, {\it ApJ} {\bf 627}, 904 (2005).
\item[25.]
M.~S.~E Roberts, C.~L. Brogan, {\it ApJ} {\bf 681}, 320 (2008).
\item[26.]
M.~S.~E Roberts, R.~W. Romani, N. Kawai {\it ApJS} {\bf 133}, 451 (2001).
\item[27.]
K.~T.~S. Brazier, G. Kanbach, A. Carrami\~nana, J. Guichard, M. Merck, {\it MNRAS} {\bf 281}, 1033 (1996).
\item[28.]
D.~A. Leahy, S. Naranan, K.~P. Singh, {\it MNRAS} {\bf 220}, 501 (1986).
\item[29.]
O. Reimer, {\it et al.}, {\it MNRAS} {\bf 324}, 772 (2001).
\item[30.]
J.~P. Halpern, F. Camilo, E.~V. Gotthelf, {\it ApJ} {\bf 668}, 1154 (2007).
\item[31.]
J.~P. Halpern, E.~V. Gotthelf, N. Mirabal, F. Camilo, {\it ApJ} {\bf 573}, L41 (2002).
\item[32.]
R. Mukherjee, J.~P. Halpern, E.~V. Gotthelf, M. Eracleous, N. Mirabal, {\it ApJ} {\bf 589}, 487 (2003).
\item[33.]
M.~C. Weisskopf, {\it et al.}, {\it ApJ} {\bf 652}, 387 (2006).
\item[34.]
J. Faherty, F.~M. Walter, J. Anderson {\it Ap\&SS} {\bf 308}, 225 (2007).
\item[35.]
G.~F. Bignami, W. Hermsen, {\it ARA\&A} {\bf 21}, 67 (1983).
\item[36.]
A.~K. Harding, J.~V. Stern, J. Dyks, M. Frackowiak, {\it ApJ} {\bf 680}, 1378 (2008).
\item[37.]
J.~P. Halpern, E.~V. Gotthelf, F. Camilo, D.~J. Helfand, S.~M. Ransom, {\it ApJ} {\bf 612}, 398 (2004).
\item[38.]
T.~M. Braje, R.~W. Romani, M.~S.~E. Roberts, N. Kawai, {\it ApJ} {\bf 565}, L91 (2002).
\item[39.]
A.~A. Abdo, {\it et al.}, ``Fermi Large Area Telescope Bright Gamma-ray Source List", {\it ApJS} submitted, arXiv {\tt 0902.1340} (2009).
\item[40.] 
The $Fermi$ LAT Collaboration is supported by the National Aeronautics and Space Administration and the Department of Energy in the United States, 
the Commissariat \`a l'Energie Atomique and the Centre National de la Recherche Scientifique/Institut National de Physique Nucl\'eaire et de Physique des Particules in France, 
the Agenzia Spaziale Italiana, the Istituto Nazionale di Fisica Nucleare, and the Istituto Nazionale di Astrofisica in Italy, the Ministry of Education, Culture, Sports, 
Science and Technology (MEXT), High Energy Accelerator Research Organization (KEK) and Japan Aerospace Exploration Agency (JAXA) in Japan, and the K.\ A.\ Wallenberg Foundation, 
the Swedish Research Council and the Swedish National Space Board in Sweden. 

Much of this work was carried out on the UCSC Astronomy Department Pleiades supercomputer. 

This work made extensive use of the ATNF pulsar database.

We thank N. Gehrels, for granting Swift/XRT time to explore the LAT error circles and the Swift MOC staff for performing the observations.

We also thank F. Camilo and S. Ransom for contributions. We thank M. Roberts for useful discussions.

\end{itemize}


\begin{table}[p]
\caption{Names and locations of the gamma-ray pulsars discovered in our blind searches. The ``Assumed Counterpart" column is the X-ray source that provided the position for the timing model, which in some cases may not be the true counterpart.\label{names_and_locations}}
{\footnotesize
\begin{tabular}{llllrrrr}
\hline\hline
{LAT PSR} & {LAT Source 0FGL} & {EGRET Source} &  {Assumed Counterpart} & {RA ($^\circ$)} & {Dec ($^\circ$)} &   {$l$ ($^\circ$)} & {$b$ ($^\circ$)} \\
\hline
J0007+7303	& J0007.4+7303 & 3EG J0010+7309		& RX J0007.0+7303 in G119.5+10.3 ({\it 37}) 	& \phn1.7565	& \phn73.0523	& \phn119.7 	& 10.5 \\
J0357+32 	& J0357.5+3205 &           		&            							& \phn59.445 	& \phn32.105 	& \phn162.7 	& -16.0 \\
J0633+0632 	& J0633.5+0634 & 3EG J0631+0642  	& Swift J063344+0632.4 						& \phn98.4333 	& \phn6.5403 	& \phn205.0 	& -1.0 \\
J1418-6058 	& J1418.8-−6058 & 3EG J1420-6038  	& R1 in G313.3+0.1  ({\it 24})			& \phn214.6779 	& \phn-60.9675 	& \phn313.3 	& 0.1 \\
J1459-60 	& J1459.4-−6056 &            		&            							& \phn224.874 	& \phn-60.878 	& \phn317.9 	& -1.8 \\
J1732-31 	& J1732.8-−3135 & 3EG J1734-3232  	&  								& \phn263.169 	& \phn-31.610 	& \phn356.2 	& 0.9 \\
J1741-2054 	& J1742.1-−2054 & 3EG J1741-2050  	& Swift J174157--2054.1  					& \phn265.4908 	& \phn-20.9033 	& \phn6.4 	& 4.9 \\
J1809-2332 	& J1809.5-−2331 & 3EG J1809-2328  	& CXOU J180950.2--233223 in G7.4-2.0 ({\it 38}) & \phn272.4592 	& \phn-23.5397 	& \phn7.4 	& -2.0 \\
J1813-1246 	& J1813.5-−1248 & GeV J1814-1228 	& Swift J181323--1246.0 					& \phn273.3475 	& \phn-12.7668 	& \phn17.2 	& 2.4 \\
J1826-1256 	& J1825.9-−1256 & 3EG J1826-1302  	& AXJ1826.1-1257 in G18.5-0.4 ({\it 26}) 	& \phn276.5356 	& \phn-12.9429 	& \phn18.5 	& -0.4 \\
J1836+5925 	& J1836.2+5924 & 3EG J1835+5918  	& RX J1836.2+5925 ({\it 30})			& \phn279.0570 	& \phn59.4250 	& \phn88.9 	& 25.0 \\
J1907+06 	& J1907.5+0602 & GeV J1907+0557 	& 								& \phn286.965 	& \phn6.022 	& \phn40.2 	& -0.9 \\
J1958+2846 	& J1958.1+2848 & 3EG J1958+2909  	& Swift J195846.1+2846.0 					& \phn299.6921 	& \phn28.7674 	& \phn65.9 	& -0.2 \\
J2021+4026 	& J2021.5+4026 & 3EG J2020+4017 	& S21 in G78.2+2.1 ({\it 33}) 			& \phn305.3777 	& \phn40.4462 	& \phn78.2 	& 2.1 \\
J2032+4127 	& J2032.2+4122 & 3EG J2033+4118  	& MT91 221 in Cygnus OB2 ({\it 32}) 		& \phn308.0612 	& \phn41.4610 	& \phn80.2 	& 1.0 \\
J2238+59 	& -- &           		&            							& \phn339.561	& \phn59.080 	& \phn106.5 	& 0.5 \\
\hline
\end{tabular}
}
\end{table}

\begin{table}[p]
\caption{Rotational ephemerides for the new pulsars. For all timing solutions, the reference epoch is MJD 54754. The given frequency and frequency derivative are barycentric. The numbers in parentheses indicate the error in the last decimal digit(s). For LAT PSR J0357+32, marked with ($\dag$), the measured frequency derivative, and thus the derived parameters, may be significantly affected by positional error (see text).We also list the number of photons, $n_{\gamma}$, obtained with the cuts used in this work (see text) over the 5 month observational period. The (1--100 GeV) photon number flux, $F_{35}$, is given, as presented in the $Fermi$ Bright Source 
List ({\it 39}), except for LAT PSR J2238+59, which was not in the Bright Source List but whose flux was derived from the same data set and analysis approach.\label{timing_solutions}} {\footnotesize
\begin{tabular}{lrrllrrr}
\hline\hline
{LAT PSR}  & {$n_{\gamma}$} & {$F_{35}$} & {$f$}  & {$\dot{f}$} & {$\tau$} & {$\dot{E}$} &  {$B$} \\ {} & {} & {($10^{-8}$  cm$^{-2}$ s$^{-1}$)} & {(Hz)} & {($-10^{-12}$ Hz   s$^{-1}$)} & {(kyr)} & {($10^{34}$ erg s$^{-1}$)} & {($10^{12}$ G)} \\
\hline
J0007+7303 	& 1509	& 6.14(27)	& 3.1658891845(5)	& \phn3.6133(3)		& 13.9	& 45.2	& 10.8 \\
J0357+32 	& 294 	& 0.64(10) 	& 2.251723430(1)	& \phn0.0610(9)$\dag$	& 585.0	& 0.5	& 2.3 \\
J0633+0632 	& 648 	& 1.60(17) 	& 3.3625440117(3)	& \phn0.8992(2)		& 59.3	& 11.9	& 4.9 \\
J1418-6058 	& 3160 	& 5.42(38) 	& 9.0440257591(8)	& \phn13.8687(5)	& 10.3	& 495.2	& 4.4 \\
J1459-60 	& 1089	& 1.26(21) 	& 9.694596648(2)	& \phn2.401(1)		& 64.0	& 91.9	& 1.6 \\
J1732-31 	& 2843 	& 3.89(33) 	& 5.087952372(2)	& \phn0.677(1)		& 120.0	& 13.6	& 2.3 \\
J1741-2054 	& 889 	& 1.31(17) 	& 2.417211371(1)	& \phn0.098(7)		& 392.1	& 0.9	& 2.7 \\
J1809-2332 	& 2606 	& 5.63(31) 	& 6.8125455291(4)	& \phn1.5975(3)		& 67.6	& 43.0	& 2.3 \\
J1813-1246 	& 1832 	& 2.79(24) 	& 20.802108713(5)  	& \phn7.615(4) 		& 43.3 	& 625.7 & 0.9 \\
J1826-1256 	& 4102 	& 2.47(27) 	& 9.0726142968(4)	& \phn9.9996(3)		& 14.4	& 358.2	& 3.7 \\ 
J1836+5925 	& 2076 	& 8.36(31) 	& 5.7715516964(9)	& \phn0.0508(6)		&1800.0	& 1.2	& 0.5 \\
J1907+06 	& 2869 	& 3.74(29) 	& 9.378101746(2)	& \phn7.682(1)		& 19.4	& 284.4	& 3.1 \\
J1958+2846 	& 1355 	& 1.29(18) 	& 3.4436636006(8)	& \phn2.6351(5)		& 20.7	& 35.8	& 8.1 \\
J2021+4026 	& 4136 	& 10.60(40) 	& 3.769079109(1)	& \phn0.7780(7)		& 76.8	& 11.6	& 3.9 \\
J2032+4127 	& 2371	& 3.07(26) 	& 6.9809351235(8)	& \phn0.9560(4)		& 115.8	& 26.3	& 1.7 \\
J2238+59 	& 811	& 0.96(11)	& 6.145017519(3)	& \phn3.722(2)		& 26.3	& 90.3	& 4.1 \\
\hline
\end{tabular}
}
\end{table}


\clearpage

\begin{figure}[htp]
  \centering
  \includegraphics[width=6in,angle=0]{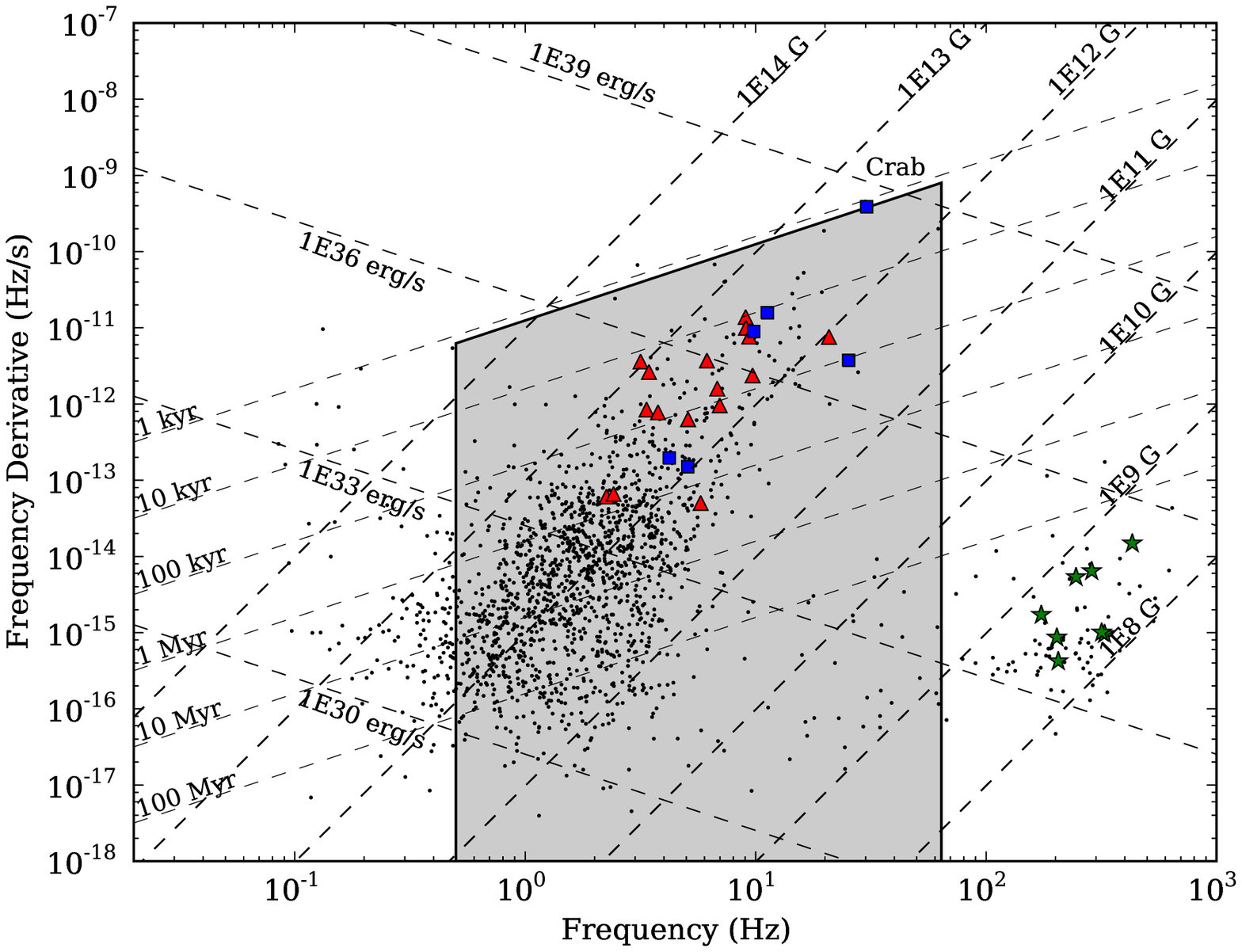}
  \label{main_plot}
\end{figure}

\noindent {\bf Fig. 1.} Frequency and frequency derivative distribution of
    the new pulsars. The dots represent the $\sim$1800 pulsars in the ATNF catalog ({\it 20}), the triangles represent the new pulsars
    reported in this paper, and the squares represent the 6
    previously-known EGRET gamma-ray pulsars, including the Crab. The stars represent the new population of gamma-ray millisecond pulsars detected by $Fermi$-LAT and 
reported in the companion publication ({\it 10}). The shaded region shows the parameter space covered by our blind search and includes 
$\sim 86\%$ of the ATNF pulsars. The three sets of lines illustrate the characteristic age $\tau = -{f}/{2\dot{f}}$, inferred surface magnetic field strength
$B = 3.2 \times 10^{19}\sqrt{-{\dot{f}}/{f^3}}$ Gauss, and spin-down power given by $\dot{E} = 4 \pi^2 I f \dot{f}$ erg s$^{-1}$, where $I$
is the neutron star moment of inertia ($I = 10^{45}$ g cm$^2$). This figure shows that the blind-search gamma-ray pulsars are drawn from the highest spin-down 
luminosity portion of the general population.
\newpage

\begin{figure}[htp]
\centering
\subfigure[J0007+7303]
{\label{Source_1}
\includegraphics[width=1.5in,angle=0]{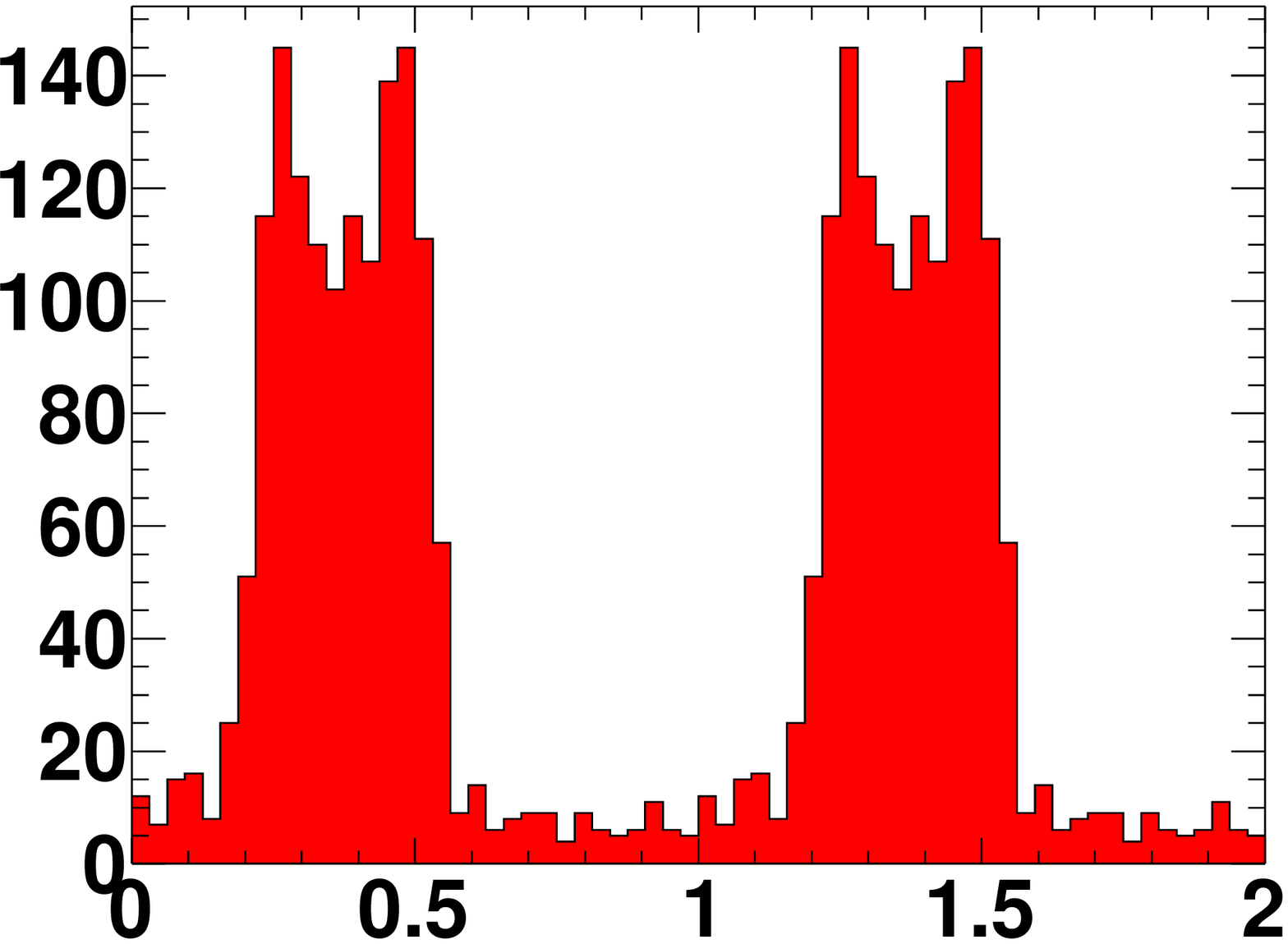}}
\subfigure[J0357+32] 
{\label{Source_2}
\includegraphics[width=1.5in,angle=0]{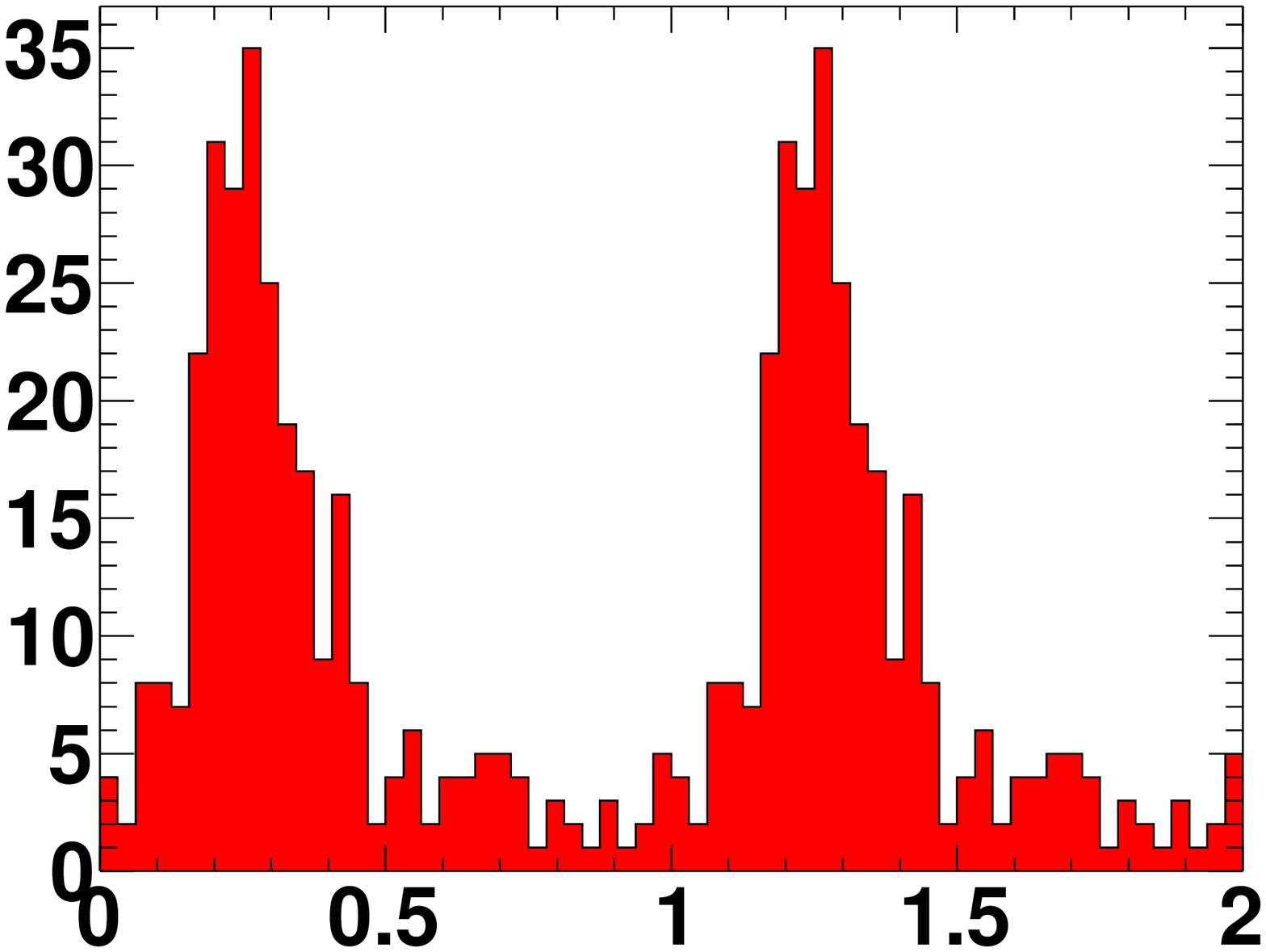}}
\subfigure[J0633+0632] 
{\label{Source_3}
\includegraphics[width=1.5in,angle=0]{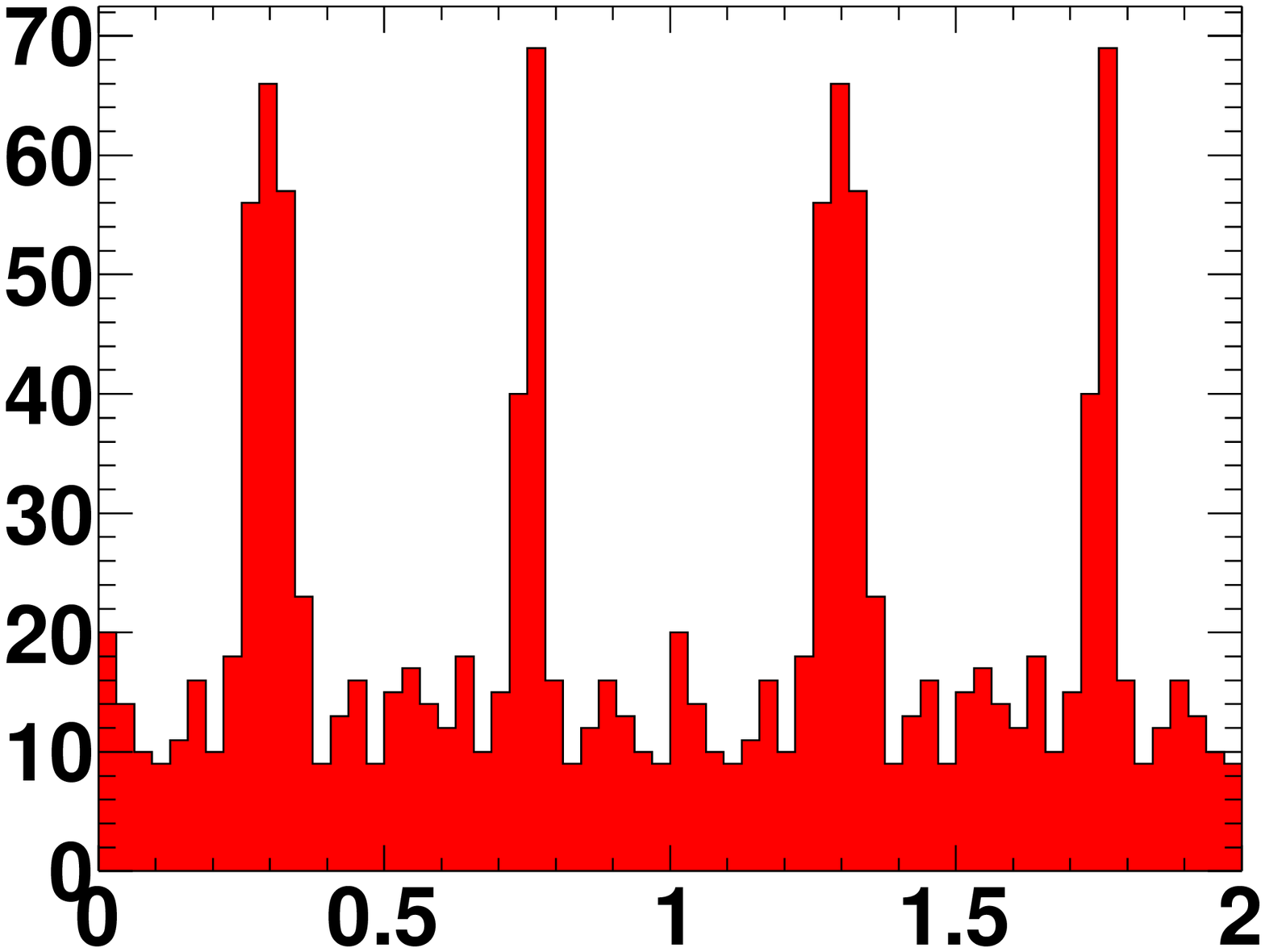}}
\subfigure[J1418-6058] 
{\label{Source_4}
\includegraphics[width=1.5in,angle=0]{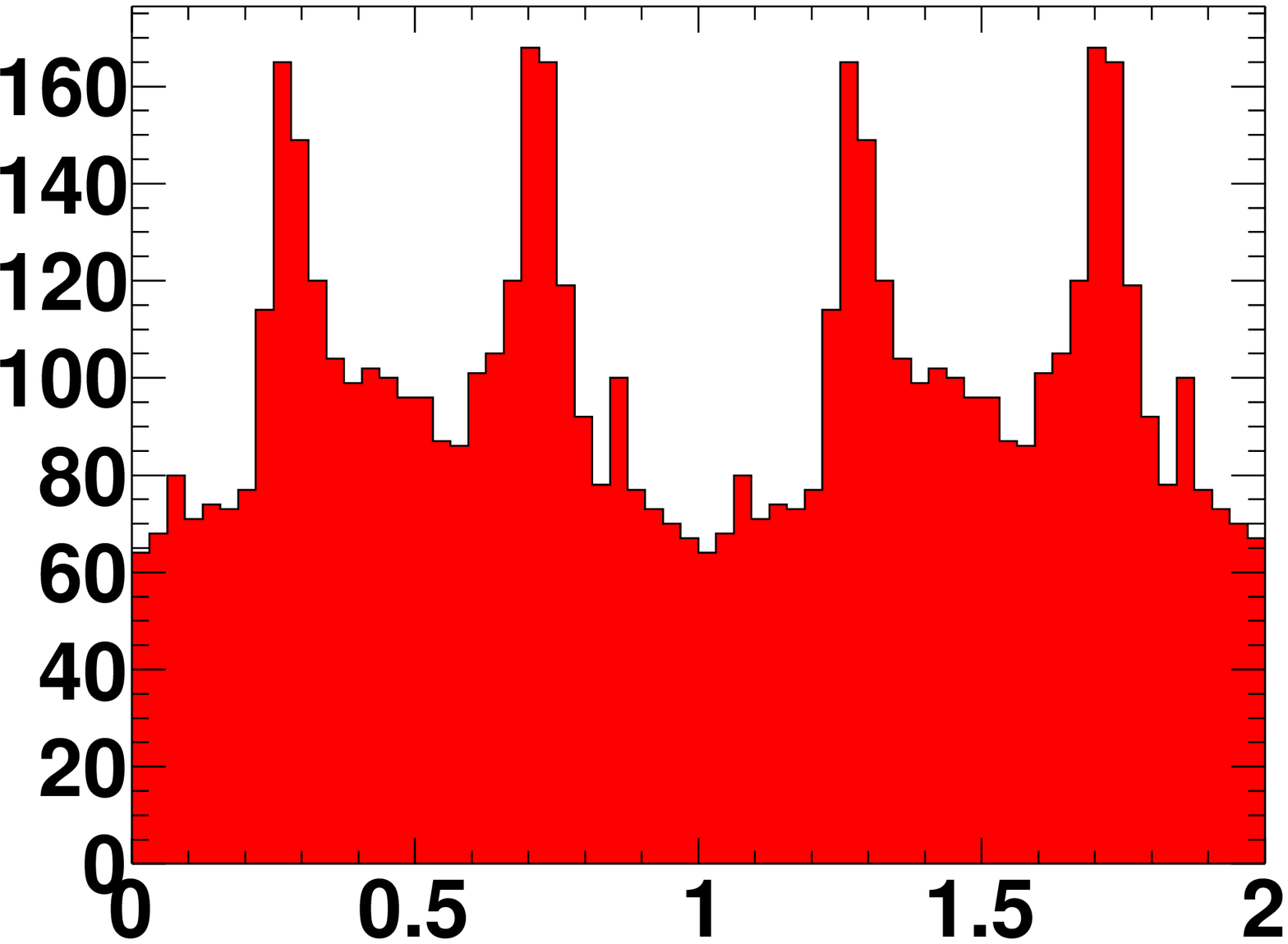}}
\subfigure[J1459-60] 
{\label{Source_5}
\includegraphics[width=1.5in,angle=0]{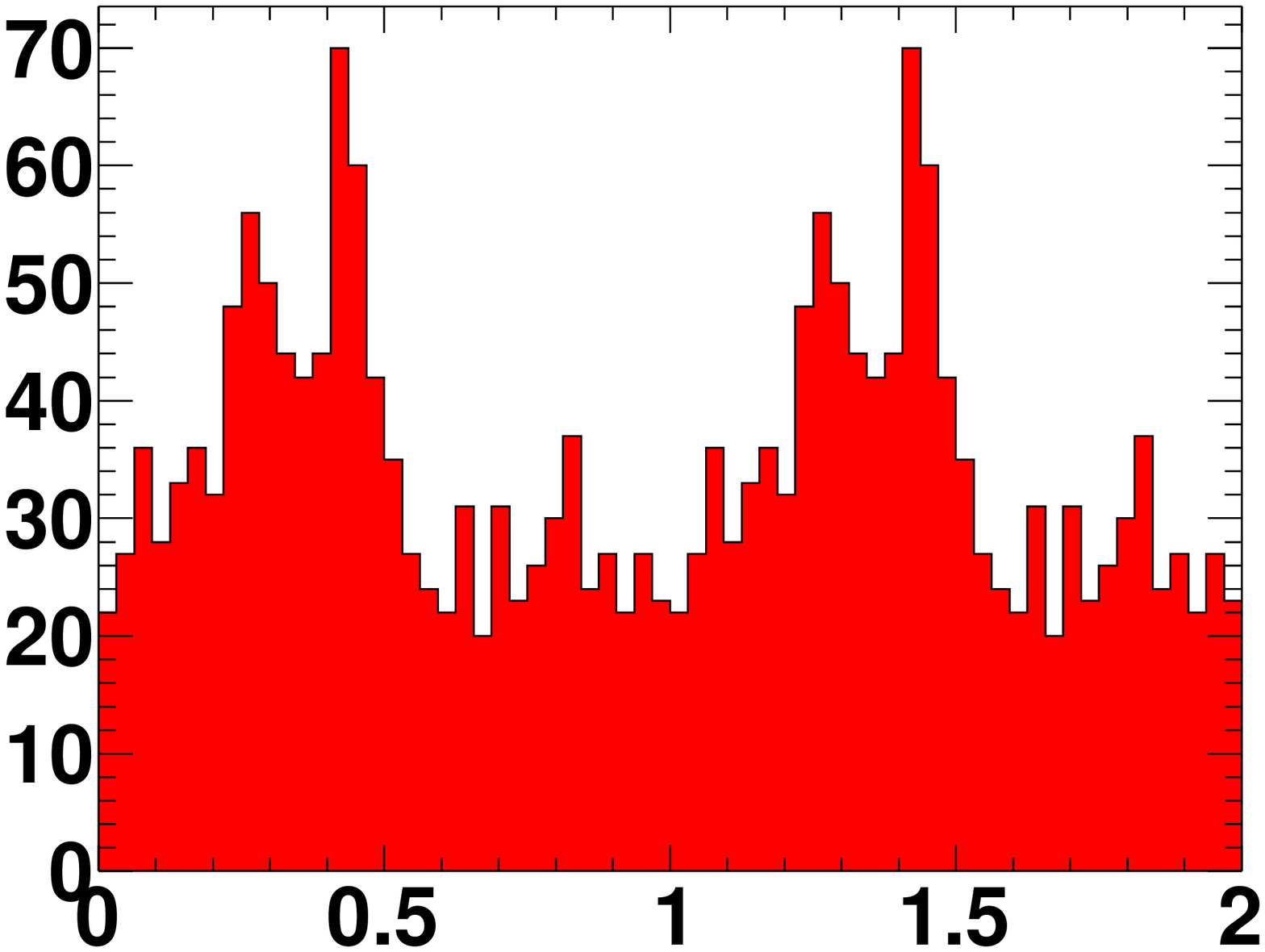}}
\subfigure[J1732-31] 
{\label{Source_6}
\includegraphics[width=1.5in,angle=0]{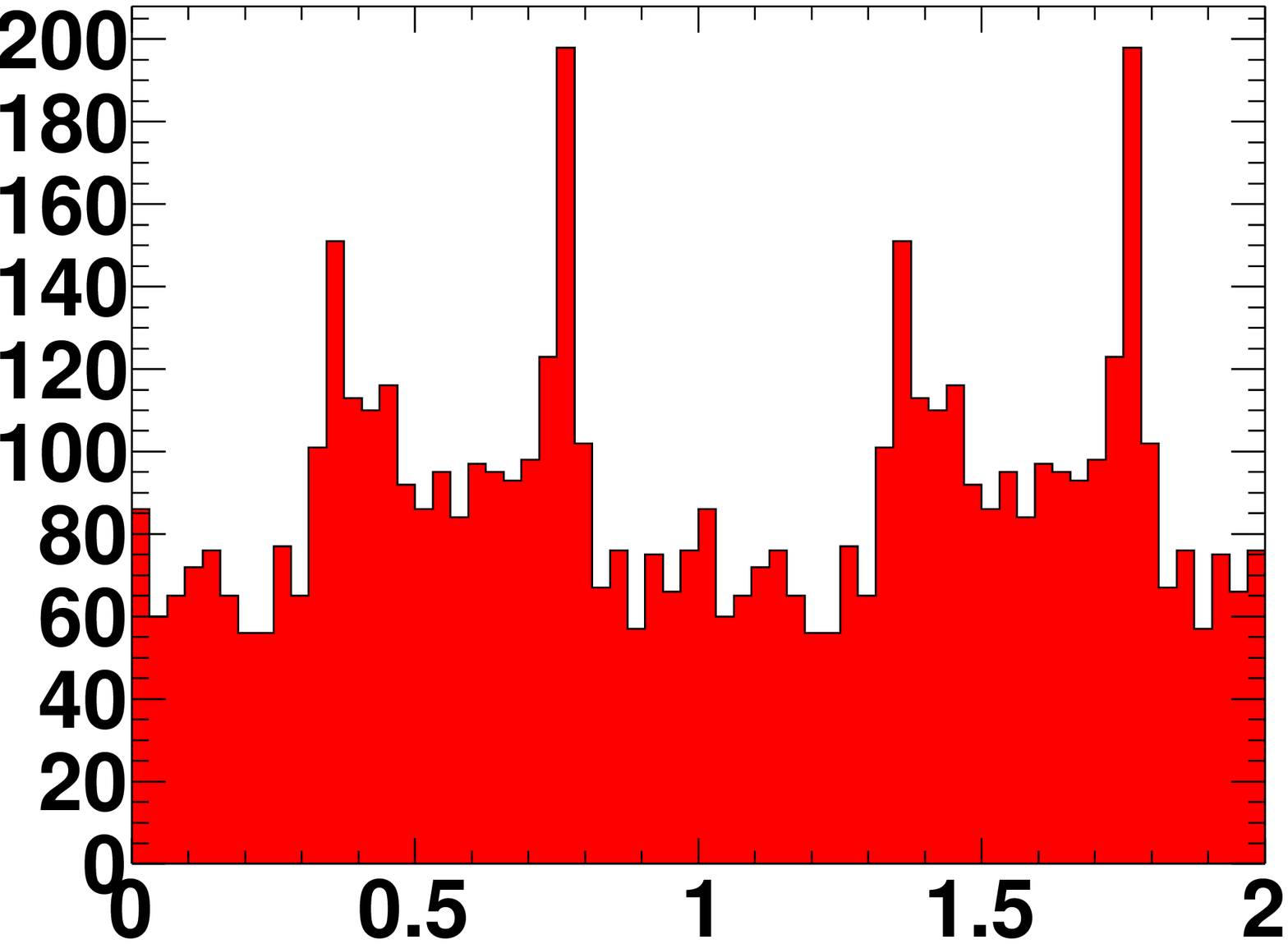}}
\subfigure[J1741-2054] 
{\label{Source_7}
\includegraphics[width=1.5in,angle=0]{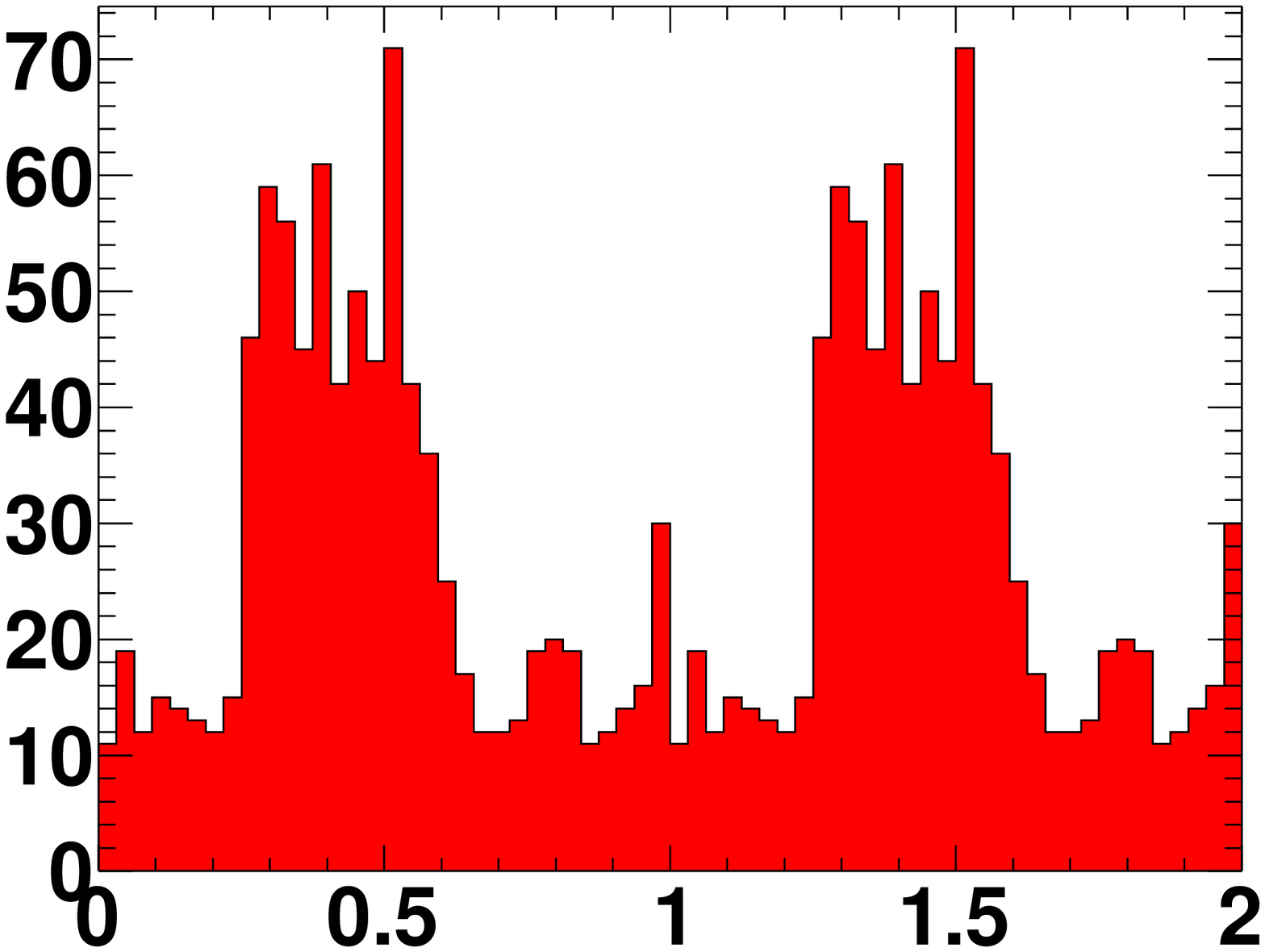}}
\subfigure[J1809-2332] 
{\label{Source_8}
\includegraphics[width=1.5in,angle=0]{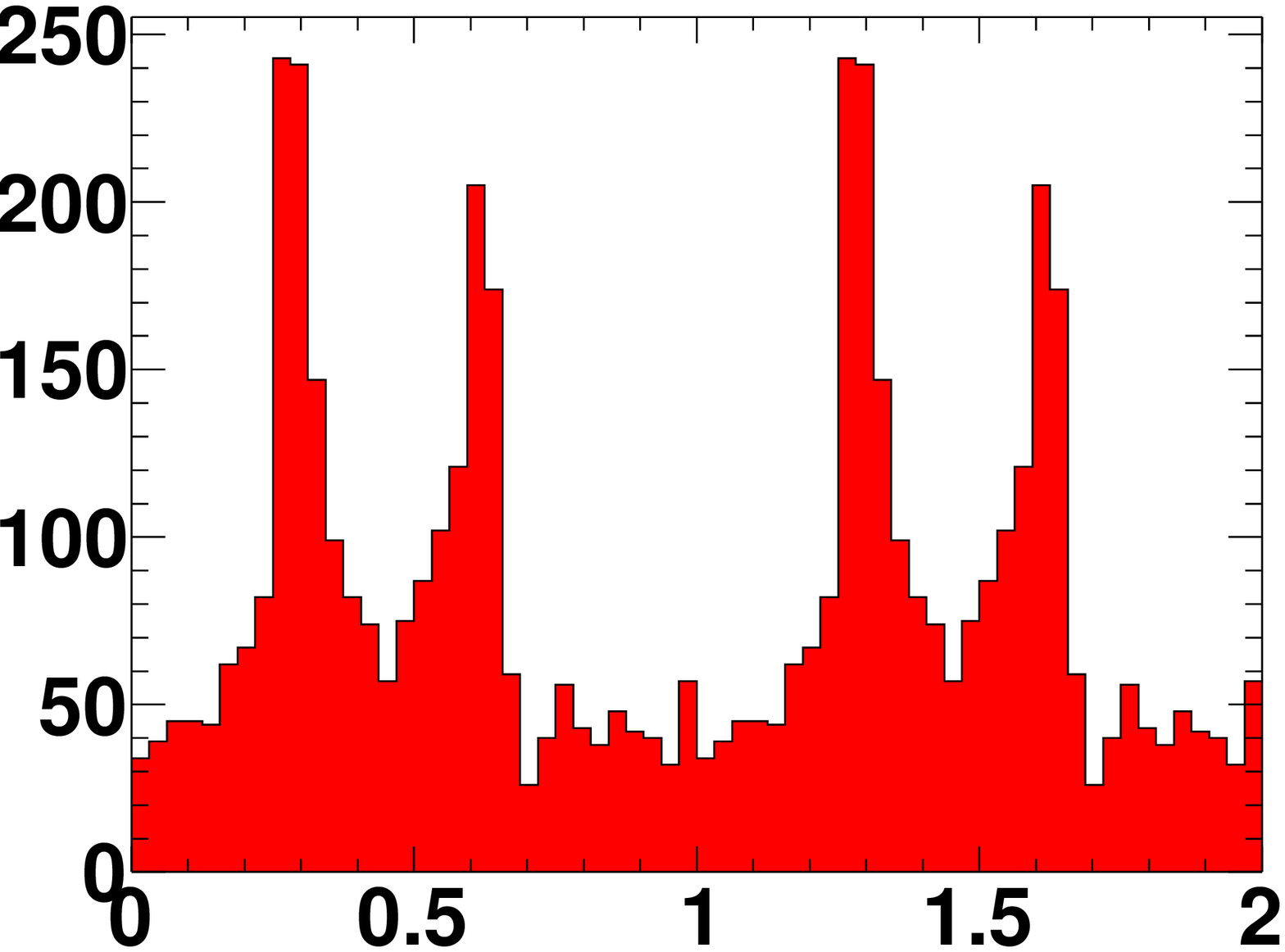}}
\subfigure[J1813-1246] 
{\label{Source_9}
\includegraphics[width=1.5in,angle=0]{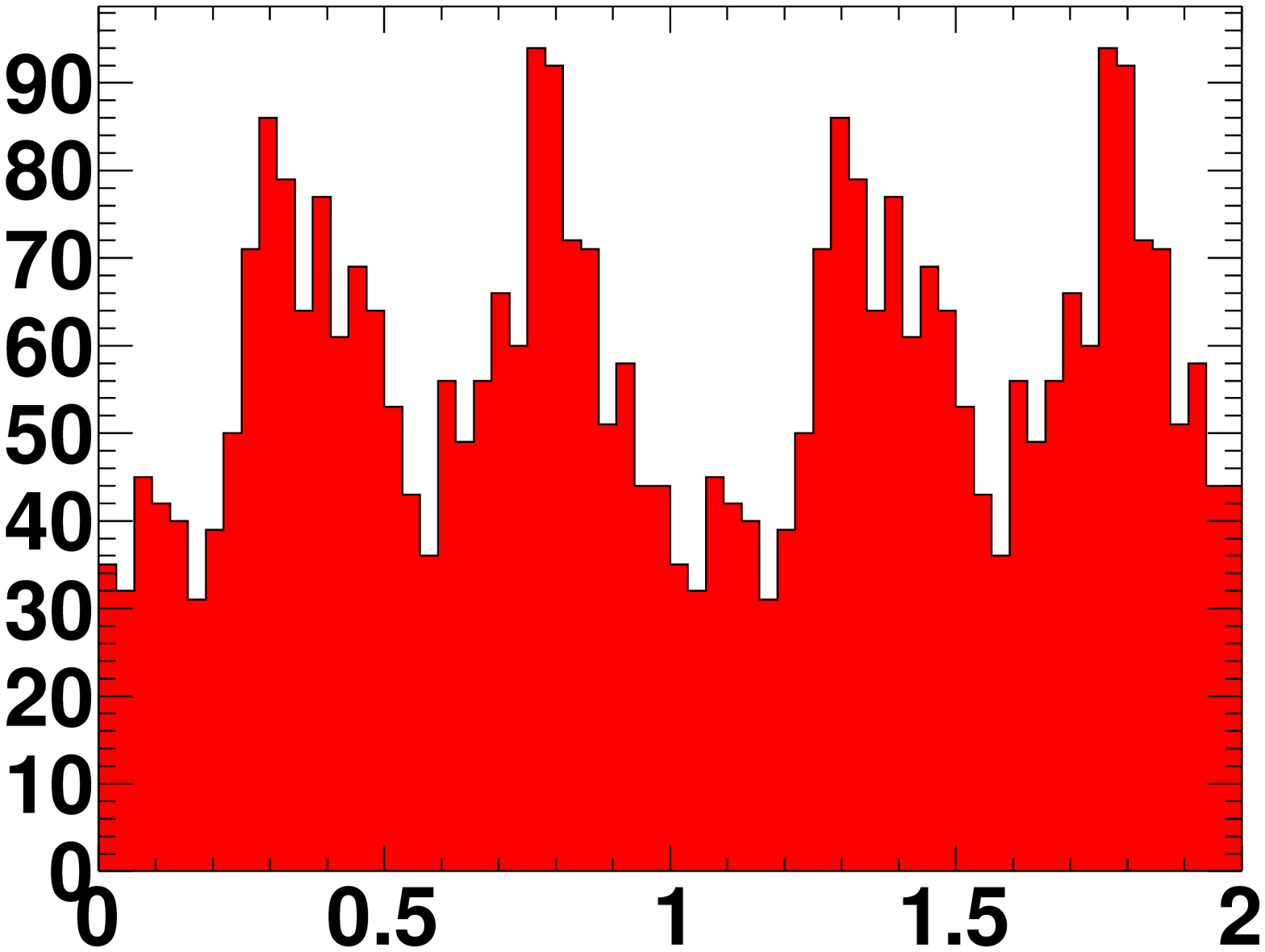}}
\subfigure[J1826-1256] 
{\label{Source_10}
\includegraphics[width=1.5in,angle=0]{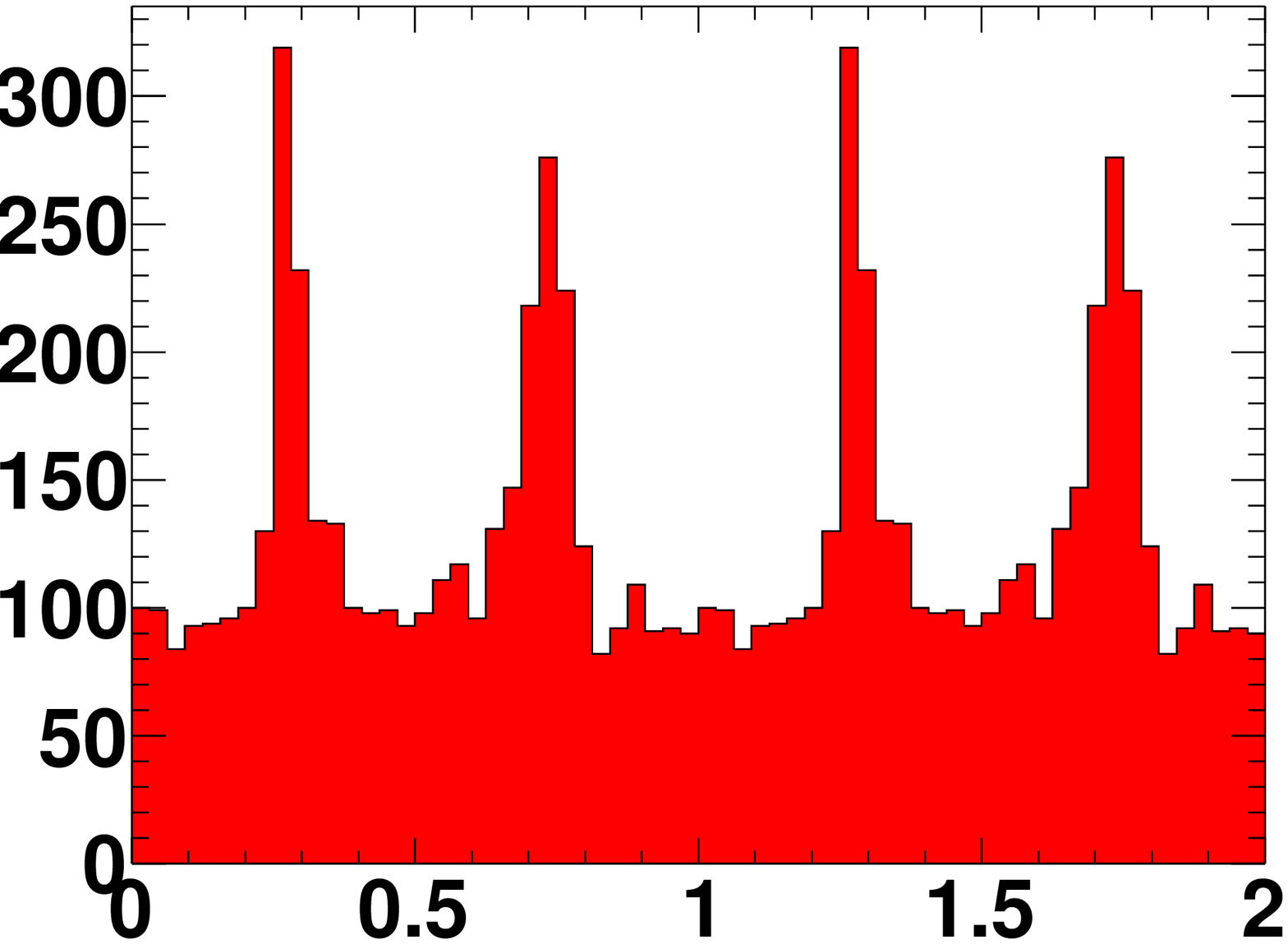}}
\subfigure[J1836+5925] 
{\label{Source_11}
\includegraphics[width=1.5in,angle=0]{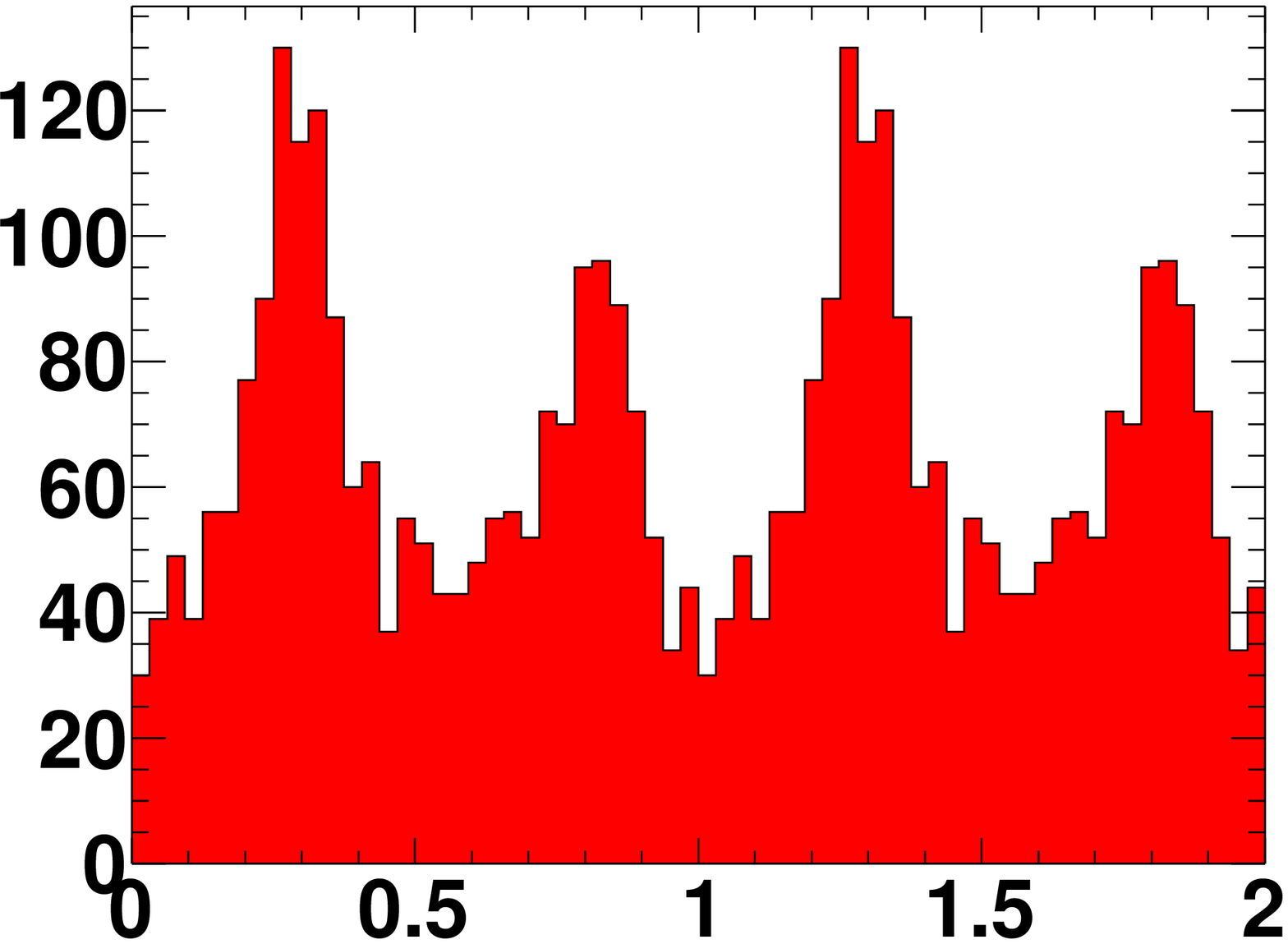}}
\subfigure[J1907+06] 
{\label{Source_12}
\includegraphics[width=1.5in,angle=0]{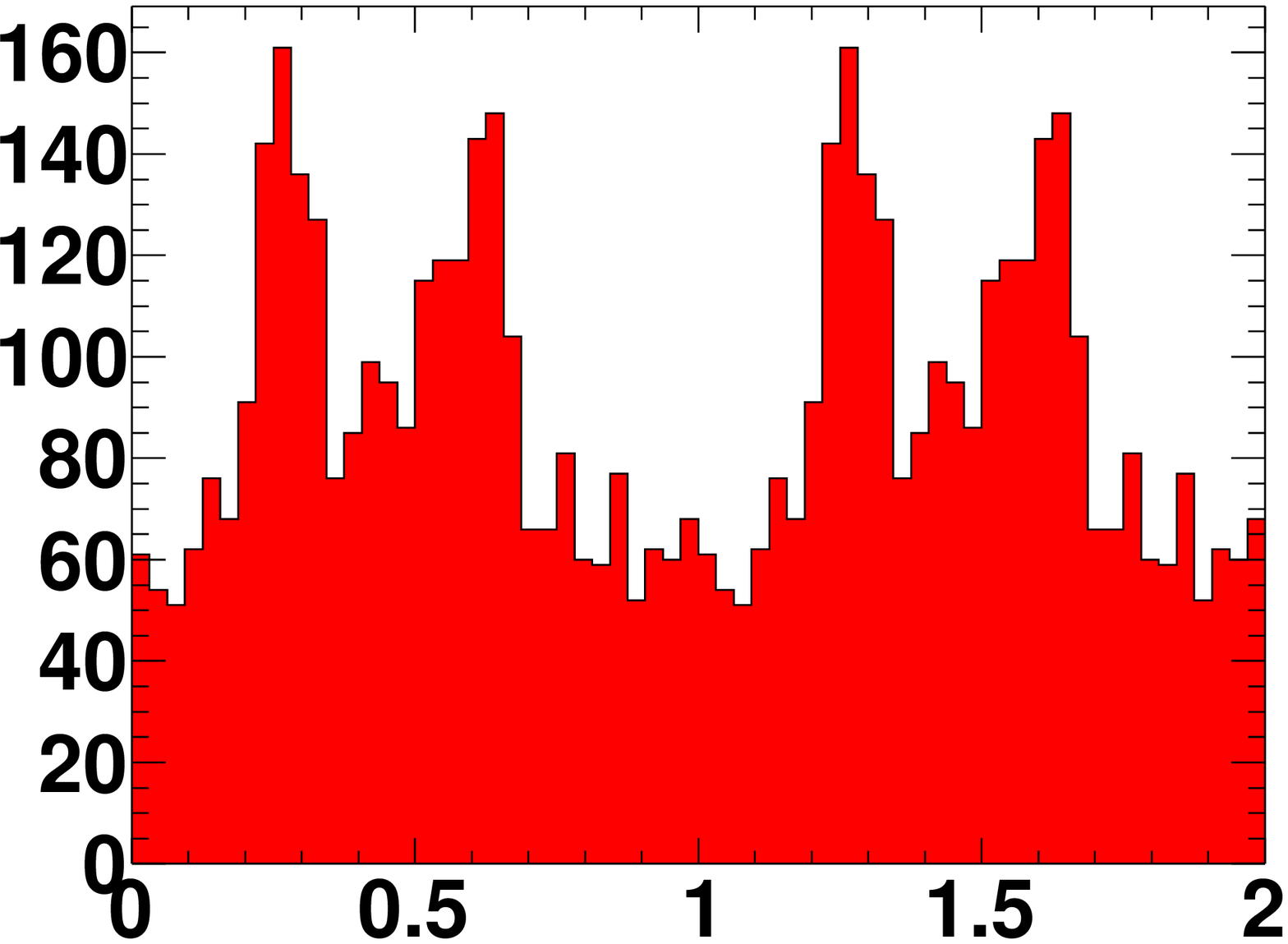}}
\subfigure[J1958+2846] 
{\label{Source_13}
\includegraphics[width=1.5in,angle=0]{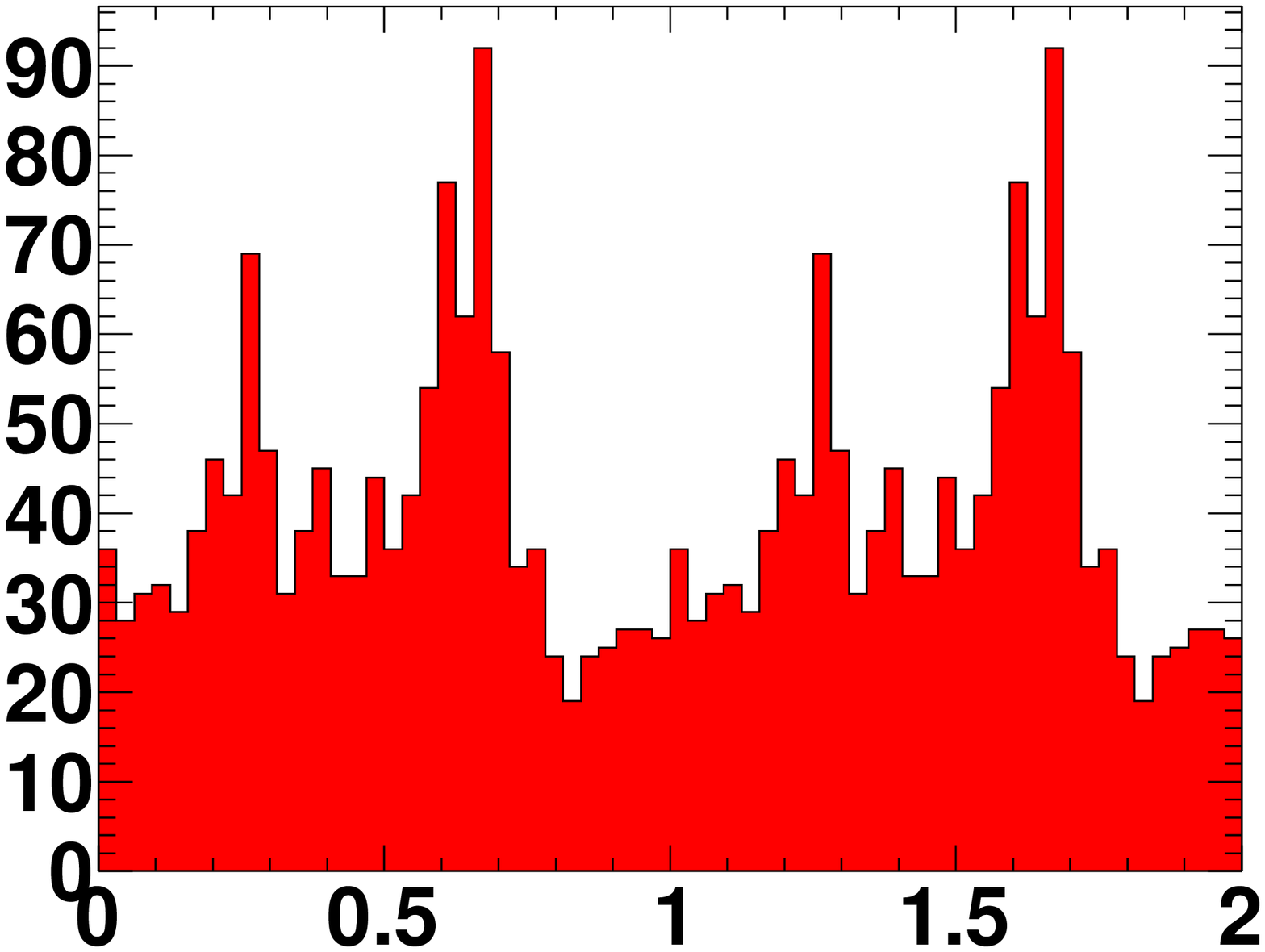}}
\subfigure[J2021+4026] 
{\label{Source_14}
\includegraphics[width=1.5in,angle=0]{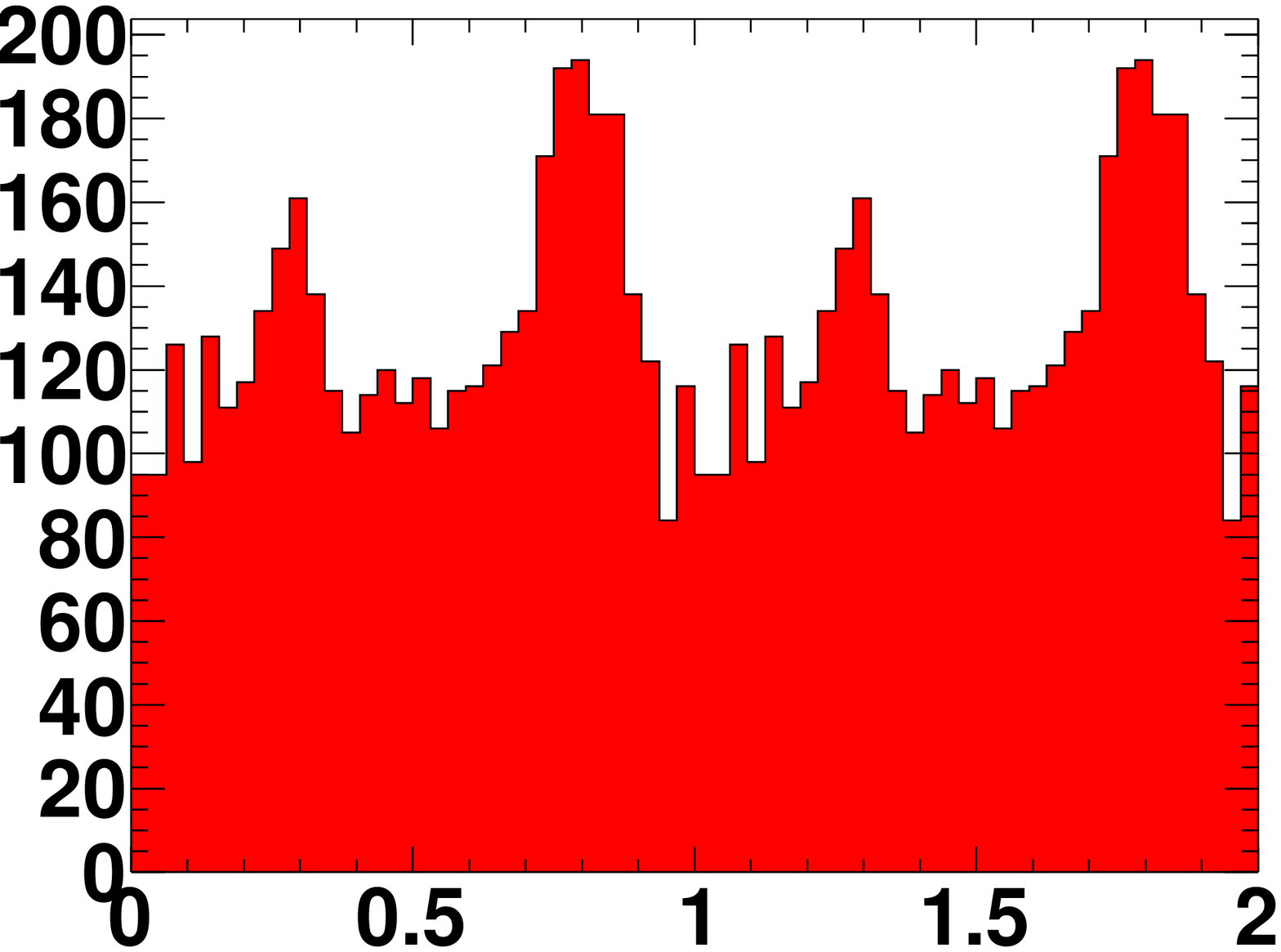}}
\subfigure[J2032+4127] 
{\label{Source_15}
\includegraphics[width=1.5in,angle=0]{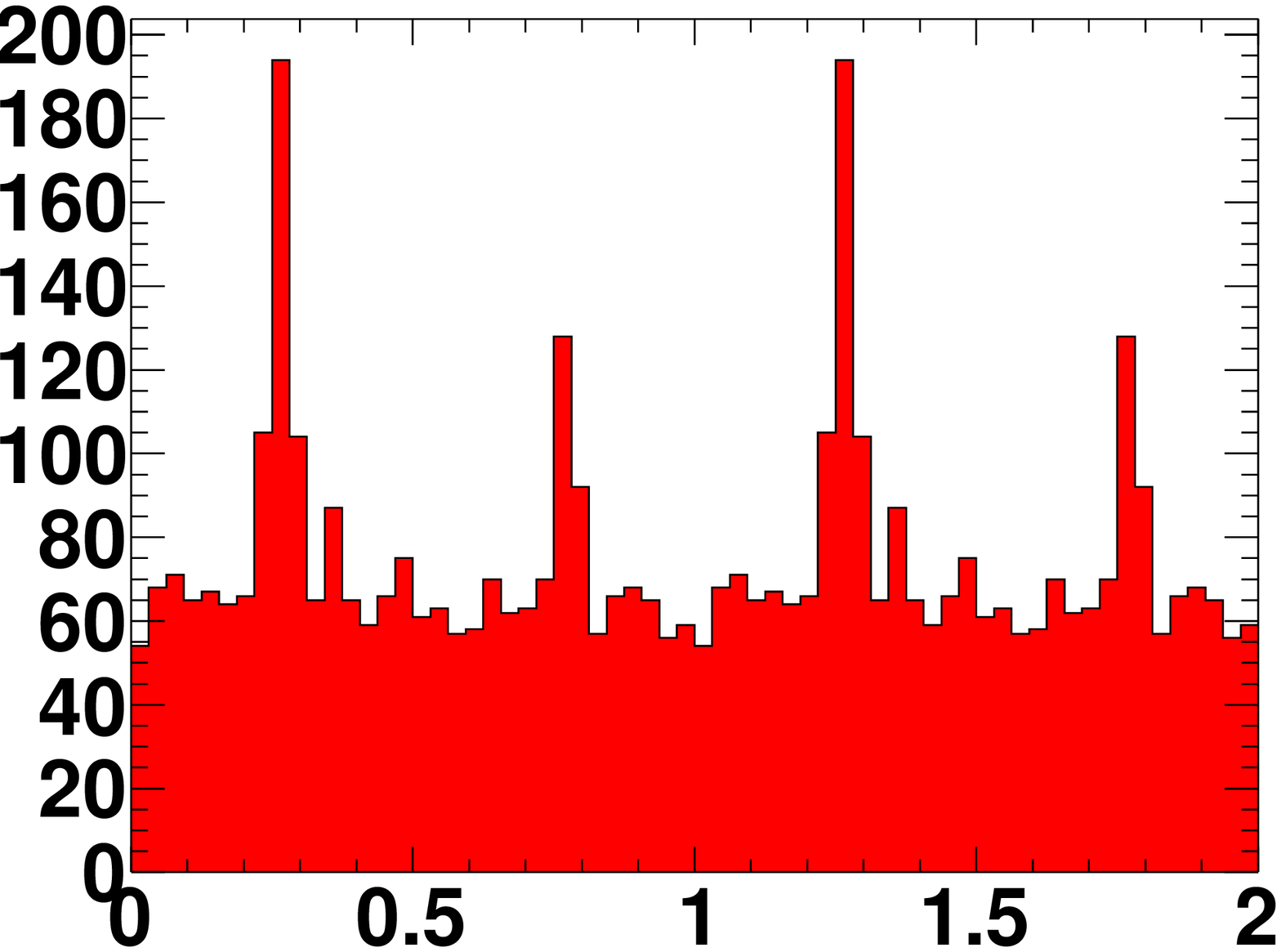}}
\subfigure[J2238+59] 
{\label{Source_16}
\includegraphics[width=1.5in,angle=0]{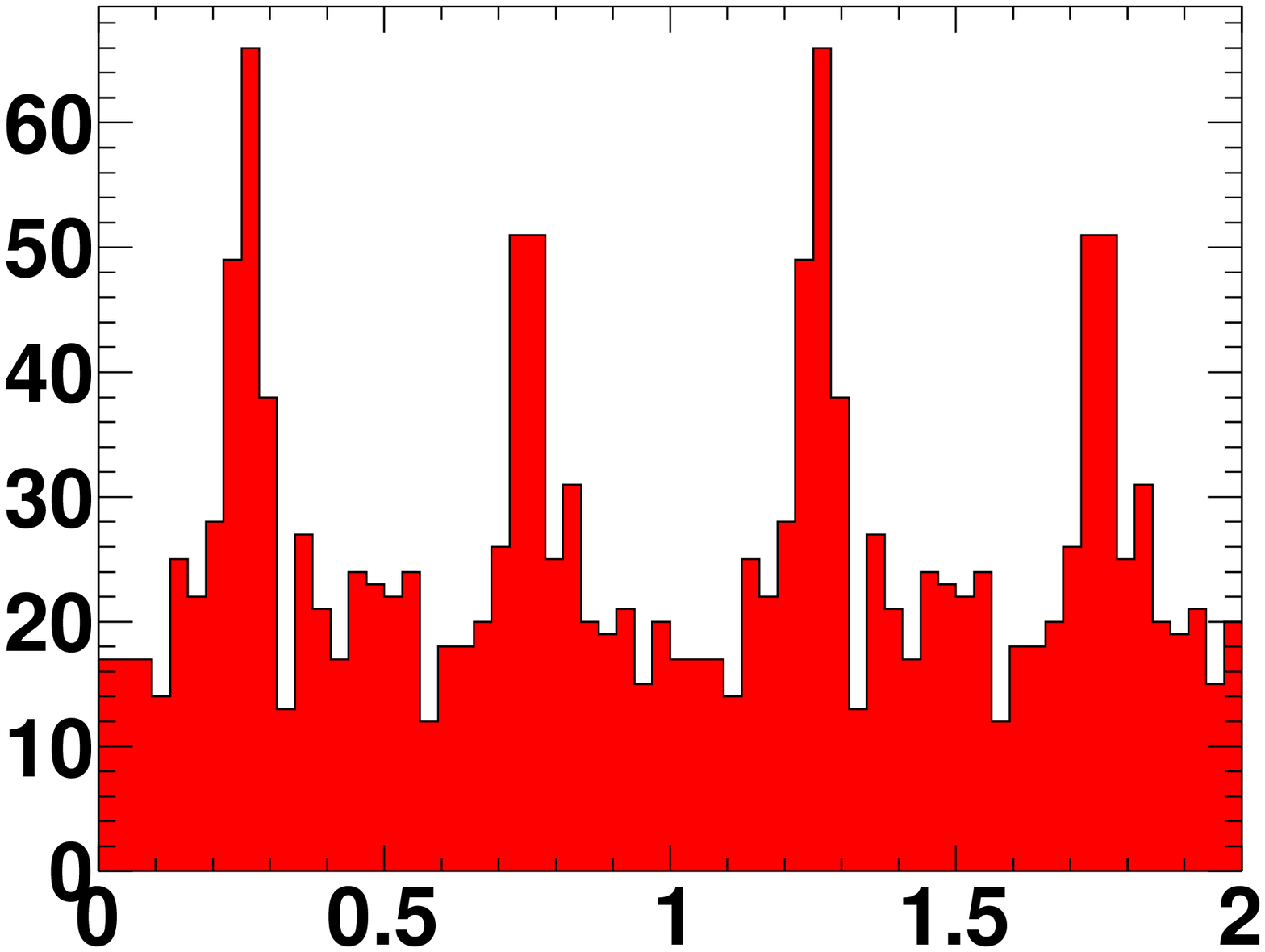}}
\label{all_lightcurves}
\end{figure}

\noindent {\bf Fig. 2.} Folded light curves, with a resolution of 32 phase bins per period, of the 16 pulsars discovered with $Fermi$-LAT, using 5 months of data with 
$E>300$ MeV, selected from a region of radius 0.8$^\circ$ around the best position for the pulsar. The light curves are not background subtracted and can include a substantial contribution from the Galactic diffuse gamma-ray emission, particularly for the pulsars at low Galactic latitude. The x axis represents phase and the y axis counts per phase bin. Two rotations are shown and the phase of the first peak has been placed at $\sim$0.3, for clarity.

\end{document}